\documentclass[twoside,a4paper]{article}
\usepackage{fleqn,epsf,graphicx}
\usepackage{authblk}
\usepackage[utf8]{inputenc} 
\usepackage{fancyhdr}
\usepackage{indentfirst}

\title{Studies of unicellular micro-organisms \textit{Saccharomyces cerevisiae} by means of Positron Annihilation Lifetime Spectroscopy}

\vspace{1cm}

\author{E.~Kubicz$^{1}$, B. Jasińska$^{3}$, B. Zgardzińska$^{3}$, T.~Bednarski$^{1}$, P.~Białas$^{1}$, E.~Czerwiński$^{1}$, A.~Gajos$^{1}$, M. Gorgol$^{3}$, D.~Kamińska$^{1}$, Ł.~Kapłon$^{1,}$ $^{2}$, A.~Kochanowski$^{4}$, G.~Korcyl$^{1}$, P.~Kowalski$^{5}$, T.~Kozik$^{1}$, W.~Krzemień$^{1}$, S.~Niedźwiecki$^{1}$, M.~Pałka$^{1}$, L.~Raczyński$^{5}$, Z.~Rajfur$^{1}$, Z.~Rudy$^{1}$, O.~Rundel$^{1}$, N.G.~Sharma$^{1}$, M.~Silarski$^{1}$, A.~Słomski$^{1}$, A.~Strzelecki$^{1}$, A.~Wieczorek$^{1,}$ $^{2}$, W.~Wiślicki$^{5}$, M.~Zieliński$^{1}$, P.~Moskal$^{1}$}

\begin{document}

\maketitle

	$^{1}$ Faculty of Physics, Astronomy and Applied Computer Science, Jagiellonian University, 
S. Łojasiewicza 11, 30-348 Kraków, Poland

       $^{2}$ Institute of Metallurgy and Materials Science of Polish Academy of Sciences, 
W. Reymonta 25, 30-059 Kraków, Poland
       
             $^{3}$ Department of Nuclear Methods, Institute of Physics, Maria Curie-Sklodowska University, Pl. M. Curie-Sklodowskiej 1, 20-031 Lublin, Poland
       
       $^{4}$ Faculty of Chemistry, Department of Chemical Technology, Jagiellonian University, 
R. Ingardena 3, 30-059 Kraków, Poland
       
            $^{5}$ Świerk Computing Centre, National Centre for Nuclear Research, 
             A. Soltana 7, 05-400 Otwock-Świerk, Poland
             
      $^{6}$ High Energy Physics Division, National Centre for Nuclear Research,
A.~ Soltana 7, 05-400 Otwock-Świerk, Poland\\

%{PACS: Put your PACS codes here}

\begin{abstract}
  Results of Positron Annihilation Lifetime Spectroscopy (PALS) and microscopic studies on simple microorganisms: brewing yeasts are presented. Lifetime of ortho - positronium (o-Ps) were found to change from 2.4 to 2.9 ns (longer lived component) for lyophilised and aqueous yeasts, respectively. Also hygroscopicity of yeasts in time was examined, allowing to check how water - the main component of the cell - affects PALS parameters, thus lifetime of o-Ps were found to change from 1.2 to 1.4 ns  (shorter lived component) for the dried yeasts. The time sufficient to hydrate the cells was found below 10 hours. In the presence of liquid water an indication of reorganization of yeast in the molecular scale was observed.
	Microscopic images of the lyophilised, dried and wet yeasts with best possible resolution were obtained using Inverted Microscopy (IM) and Environmental Scanning Electron Microscopy (ESEM) methods. As a result visible changes to the surface of the cell membrane were observed in ESEM images.
\end{abstract}


%
%  The body of the paper starts here
%

\section{Introduction}

Positron and its bound state positronium are applied since a few decades as a tool in material investigations.  Especially PALS method is widely  used, as it gives a possibility to correlate the mean o-Ps lifetime value with cavity size in which annihilation undergoes. The relationship called Tao-Eldrup model \cite{4, 24} describing o-Ps pick-off annihilation, assumes free space as a spherical potential well in which positronium is trapped. It allows to find the dependence of the mean o-Ps lifetime value in the trap and its radius. Some modifications concerning other shapes and larger sizes \cite{5,6,7,10,11} widespread  the application area for new classes of media like for example porous materials containing elongated pores, sometime opened to the surface.   
	PALS allows to study many properties of investigated materials like presence of defects, thermal expansion, temperature of phase transition in polymers, processes of gases or steams sorption in pores. However, it was applied in a very limited number of papers concerning living biological material \cite{1,20}. Some applications in biological system were described by Jean and Ache in 1977 \cite{12} and were further studied in Jean's group. They focused on studies of healthy and abnormal skin samples \cite{13, 15} and reported that S parameter from Doppler broadening technique is correlated with broadly defined level of skin damage. Over last few years biological systems appeared in interest of annihilation techniques again \cite{2}. The precise investigations seem to be very complex because of presence of water in which positronium also undergoes annihilation. However, hydrated solid materials were studied using PALS \cite{23} successfully. In the paper by Hugenschmidt et al. \cite{9} some experiments concerning behaviour of the free volume on $ H_{2}O $ loading and uniaxial pressure on glucose - gelatin compounds were performed. 
	In this study we have performed a test experiment showing the possibility to observe dynamics of the water sorption by hydrophilic material - lyophilised yeast cells. 
	Yeasts are eukaryotic micro-organisms, a spherical unicellular microscopic fungus.  They are living micro-organisms in both the presence or absence of oxygen in the environment. For presented studies \textit{Saccharomyces cerevisiae} were used. These are most commonly used for brewing, as they ferment by converting carbohydrates to a carbon dioxide and alcohols. The diameter of a single yeast cell amounts of average 4 – 6 $ \mu m $. These species of yeasts is also a commonly used as a model organism in cell biology research \cite{3,8}.

\section{Experimental}
\subsection{Sample preparation}

Lyophilised \textit{Saccharomyces cerevisiae} cells used in these measurements are a commercial product Instaferm Instant Yeasts manufactured by Lallemmand Iberia, SA. Used in the distilling industry. Product "Instaferm" has the form of fine granules, color beige with size of about 1.5 $ \times $ 0.5 [mm] with less than 7\% of liquids and 1 g of lyophilised yeasts containing min. $ 3*10^{9} $ live cells. In 10\% solution pH of yeasts is of about 6.
	For microscopic and PALS studies yeasts were used in three forms as 1) lyophilised cells (as originally produced by manufacturer)  2) dried, after addition of water to lyophilised cells, rinsing and evaporating in standard conditions and 3) wet with added drop of water.

\subsection{Experimental techniques}

	A standard "fast-slow" delayed coincidence PAL spectrometer was used to conduct  measurements. The $ ^{22}$Na positron source of the activity of about 0.4~MBq, surrounded by two layers of sample, was placed inside a chamber. The $ \gamma $ quanta were registered using two detectors equipped with cylindrical BaF$ _{2} $ scintillators of sizes $ \phi $ 1.5" $ \times $ 1.5" each.  All measurements were conducted in room temperature in four stages: 1) at the beginning material was investigated  in vacuum in order to remove small amounts of water possibly present in the yeast, 2) next  dried air was introduced to the sample chamber, as sorption experiment was planned to perform under normal pressure, 3) with presence of water vapour (wet filter paper was placed  in the chamber), and 4) with drop of water placed directly to the yeast. Spectra with a total statistics of  $\sim 10^{6}$ counts for each experimental point were analyzed using LT \cite{14} and MELT programs (for integrated ones with statistics of $\sim 10^{7}$).
	ESEM (Environmental SEM) images were obtained using "Quanta 3D FEG" microscope, produced by FEI company working in environmental mode, dedicated as Gaseous Secondary Electron Detector (GSED). Measurements were conducted in the room temperature, under pressure   of water vapour of 130 Pa and a beam with energy of about 30 keV.

\section{Results}

Positron lifetime spectra of yeasts samples were measured as a function of hydration time. The intensity of o-Ps component is commonly accepted as the source of information about concentration of free volumes in the medium, where the o-Ps lifetime is measured as a value of free volumes sizes.  In the PAL spectra measured in lyophilised material, two components related to o-Ps annihilation in the voids were found. The first one with mean lifetime of about $ \tau_{3} \simeq$1.3 ns and the second one $ \tau_{4} \simeq$ 2.9 ns. Free volumes radii determined from Tao-Eldrup model assuming spherical shape of void, reach about $ r_{3} \simeq$  0.2 nm and $ r_{4} \simeq$  0.35 nm respectively. In Fig. 1 the changes of the o-Ps lifetimes ($ \tau_{3} $ , $ \tau_{4} $) and intensities ($ I_{3} $, $ I_{4} $) as a function of time and sample environment are shown. 

\begin{figure}[h!]
\begin{center}
\includegraphics [scale=0.46] {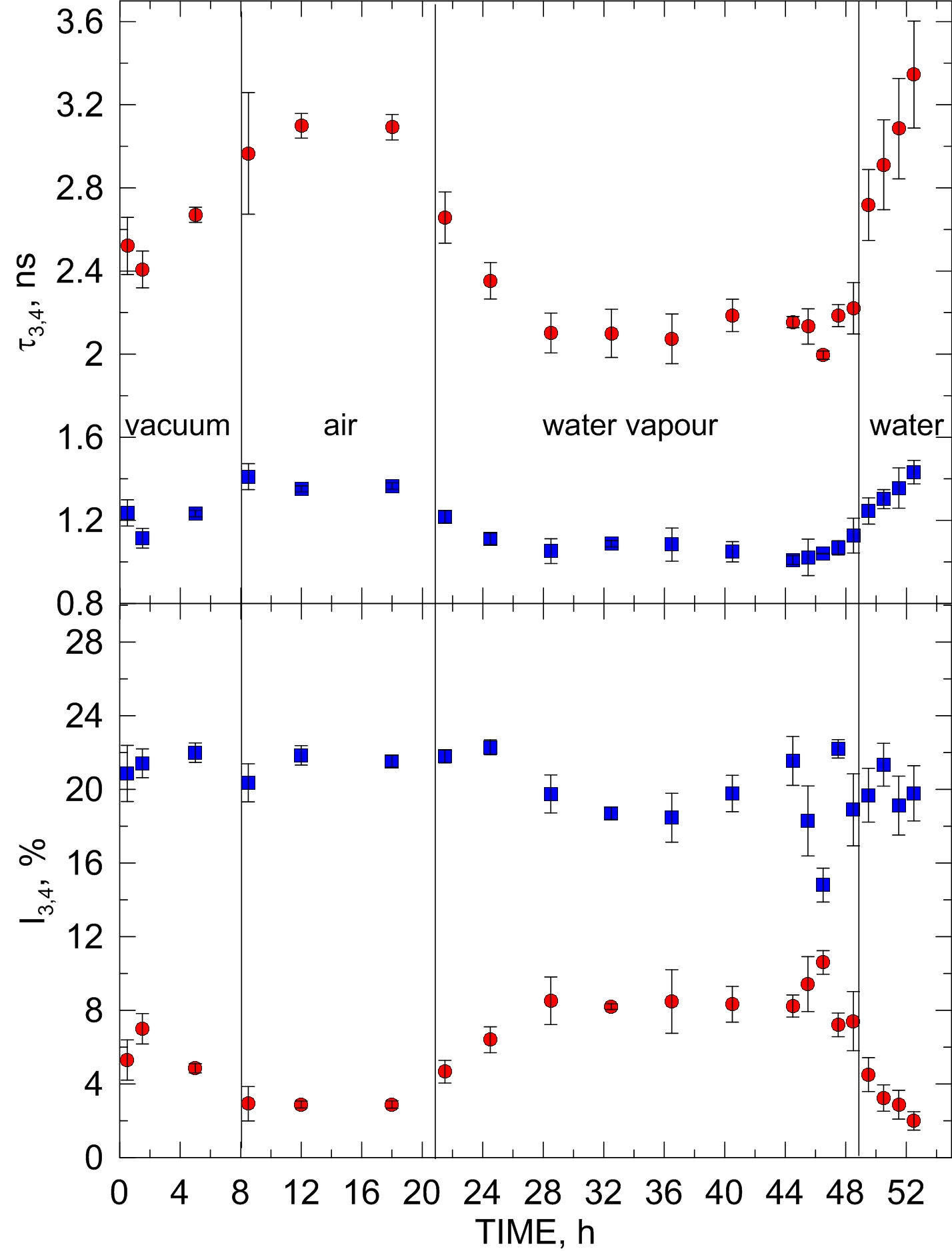}
\caption{The o-Ps lifetime and intensity values as a function of the water vapour sorption time, index 3 denotes shorter-lived component (squares) while 4 – longer-lived (circles). Measurement were conducted in four stages: 1) in vacuum , 2) in dried air, 3) with presence of water vapour, and 4) with drop of water placed in the chamber containing yeast.}
\end{center}
\end{figure}

\begin{figure}[h!]
\begin{center}
\includegraphics [scale=0.45] {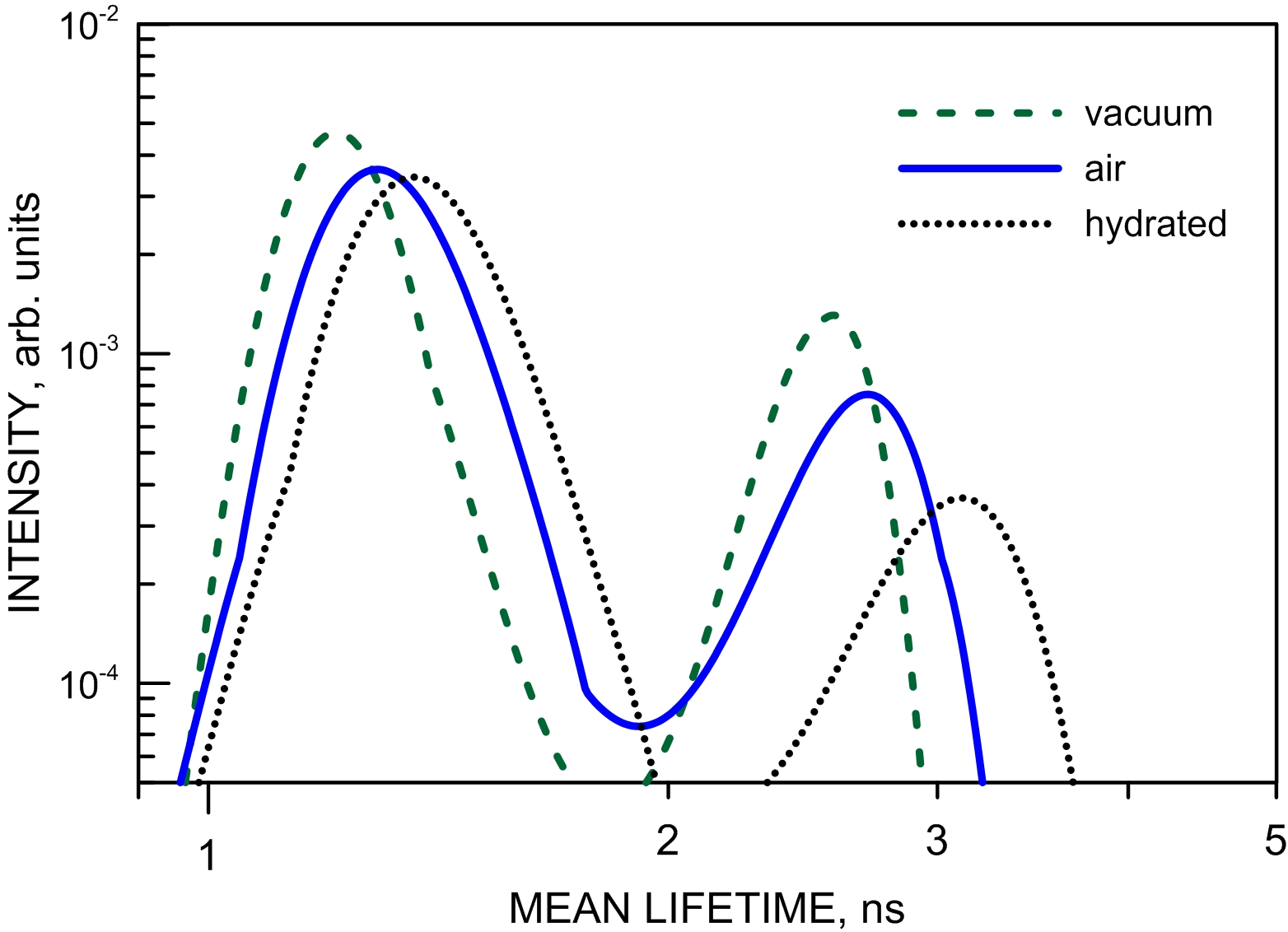}
\caption{Distributions of the o-Ps lifetimes in lyophilised yeasts in vaccum, than with the chamber open (air) and after adding drop of water to the chamber (hydrated).}
\end{center}
\end{figure}

 In the course of water vapour sorption, the lifetime value of the fourth component decreases to about 2.0 ns and its intensity value rises to about 8\%, whereas changes of parameters describing $ \tau_{3} $ component are much smaller. It is a result of water loaded to the material via larger voids having probably shape of elongated channels open to the surface. The o-Ps lifetime value (in presence of water vapour) of a range of 2.0 ns is known from literature as o-Ps lifetime value in aqueous solutions. In the presence of liquid water (4th stage) both o-Ps lifetime values rapidly rise. This may be a result of water molecules entering the cell by osmosis process and possible changes in cell's wall proteins conformation, but since yeasts cells have never before been studies at such small scale (below 1 nm) explanation of these results needs additional investigation. One can only say that it suggest reorganization of material in the molecular scale. 
\indent Additionally in Fig. 2. distributions of the o-Ps lifetimes in lyophilised and hydrated yeast are presented, as one can  observe mean lifetimes of both o-Ps components are longer for cells with addition of water, these probably  is the result of channels in the cell wall being elongated due to the present of water, but further investigation is needed on this subject.
 
\vspace{0.5cm}
 
\indent In order to see how surface of yeasts cells changes in presence of water, studies with IM 
(Fig. 3.) and  ESEM (Fig. 4) were performed. In Fig. 3. it  is well visible that after adding drop of water (upper right) cells starts to grow and move on the glass filling all voids.

\begin{figure}[h!]
\begin{center}
\includegraphics [scale=0.35] {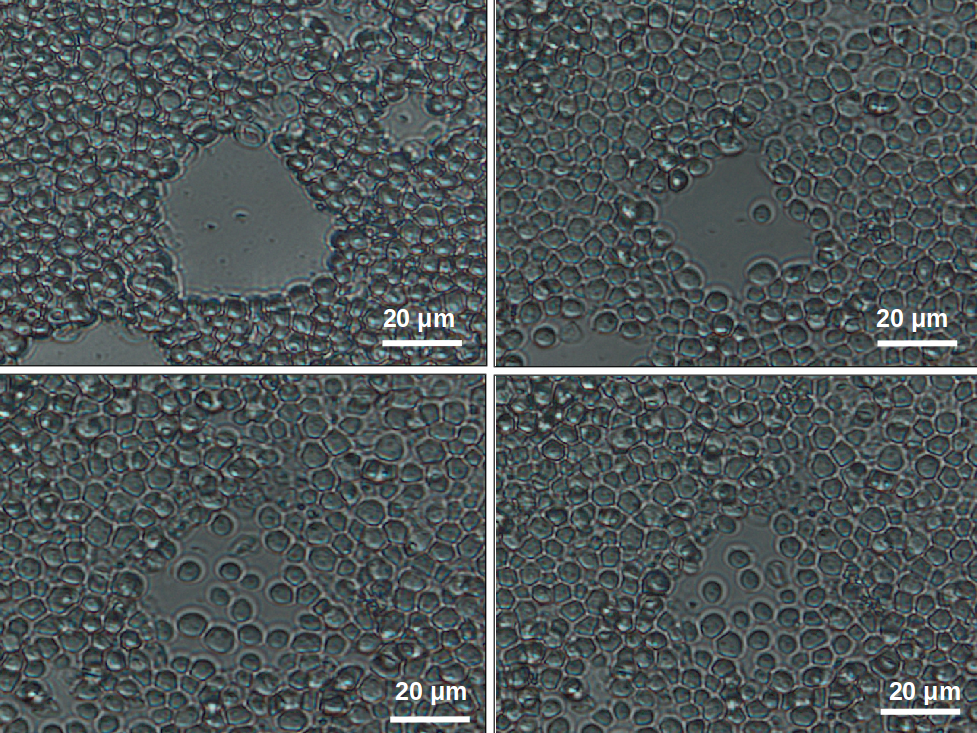}
\caption{Inverted microscopy images for dried yeasts (upper left), 1 second after adding drop of water (upper right),  1 min after (bottom left), and 5 min after (bottom right).}
\end{center}
\end{figure}

\begin{figure}[h!]
\begin{center}
\includegraphics [scale=0.34] {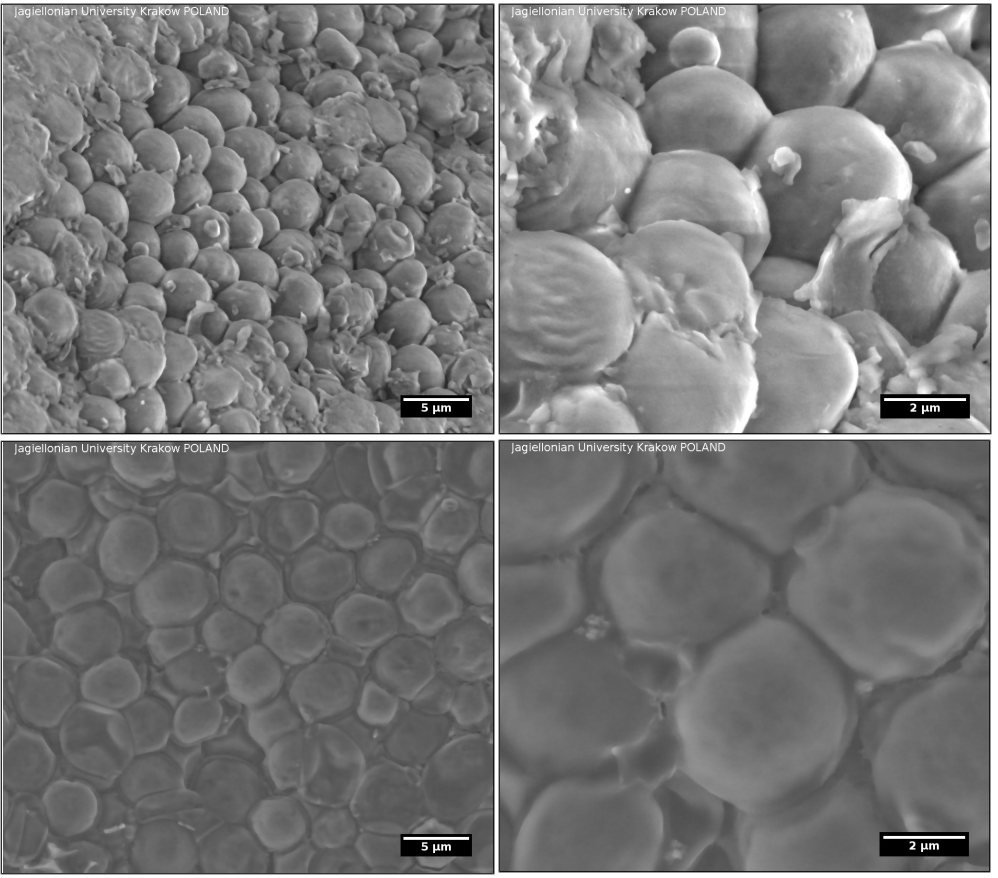}
\caption{Environmental Scanning Electron Microscopy  images of  lyophilised yeasts (upper) and  dried under normal conditions, after addition of water (bottom).}
\end{center}
\end{figure}

\indent  In Fig. 4. results from ESEM are presented. As one can observe, this method show that cell surface structure is clearly changed after addition of water, not only cells  growths, but its surface is smother and shape of cells are also changed. This is in line with the result inferred from  PALS analysis,  that some changes on molecular level are occurring in yeasts during hydration process. 
 
\newpage
\section{Conclusions}

In this study an experiment showing the possibility to observe dynamics of the water sorption by a hydrophilic material like lyophilised yeast cells was performed. As a result we proved that PALS can be successfully used for studies of biological materials in the nanoscale. Knowing the mean lifetimes of both o-Ps components, we were able to evaluate sizes of free volumes in yeasts, which in presented results were smaller than 1 nm.
 	The microscopic data allows to observe the cell and its membrane in the scale of micrometers, while from PALS measurements we received information in the scale below 1 nm. We observed changes of free volumes between molecules in cell membrane. As the single yeasts cells are of range of 2 – 6 $ \mu m $, the space between the cells is of the same range and is probably seen by PALS as a vacuum level. 
	In the future these studies of model biological structures like yeasts, will enable determination of early and advanced stages of carcinogenesis by observing changes in biomechanical parameters between normal and tumor cells employing PALS method combined with J-PET tomography system \cite{16} which is currently being developed at Jagiellonian University \cite{17,18,19,21,22,25}.

\section{Acknowledgements}

We acknowledge technical and administrative support by T. Gucwa-Ryś, A. ~Heczko, M. Kajetanowicz, G. Konopka-Cupiał, W. Migdał and the financial support by the Polish National Center for Development and Research through grant INNOTECH - K1/1N1/64/159174/NCBR/12, the Foundation for Polish Science through MPD programme and the EU and MSHE Grant No. POIG.02.03.00 - 161 00-013/09. We wauld also like to acknowledge Dr M. Targosz – Korecka and Dr B. R. Jany for help with microscopic measurements and images interpretation. Microscopy research was carried out with the equipment purchased thanks to the financial support of the European Regional Development Fund in the framework of the Polish Innovation Economy Operational Program (contract no. POIG.02.01.00-12-023/08) and supported by 1.1.2 PO IG EU project POMOST FNP: “Elasticity parameter and strength of cell to cell interaction as a new marker of endothelial cell dysfunction in hyperglycemia/hypoglycemia”

%  The bibliography starts here
%  The first three items serve as the examples for
%  the citation of paper, book and article in proceedings.
%


\begin{thebibliography}{}

\bibitem{1}  Al-Sheibani Z. T. et al., (2012) Structural and electrical properties of CdO/porous-Si heterojunction,  Iraqi Journal of Physics 10, 77.

\bibitem{2} Axpe E et al., (2014) Detection of Atomic Scale Changes in the Free Volume Void Size of Three-Dimensional Colorectal Cancer Cell Culture Using Positron Annihilation Lifetime Spectroscopy, PLoS One 2;9(1). DOI: 10.1371/journal.pone.0083838 

\bibitem{3}Barnett J. A. , Payne R. W. , Yarrow D., (1983) Yeasts: characteristics and identification, Cambridge University Press

\bibitem{4}  Eldrup M, Lightbody D, Sherwood JN (1981) The temperature dependence of positron lifetimes in solid pivalic acid. Chem. Phys. 63, 51-58. DOI: 10.1016/0301-0104(81)80307-2  

\bibitem{5} Gidley D. W.  et al., (1999) Positronium annihilation in mesoporous thin films, Phys. Rev. B 60, R5157(R). DOI: 10.1103/PhysRevB.60.R5157

\bibitem{6} Goworek, T. Ciesielski, K. Jasinska, B. Wawryszczuk, J., (1997) Positronium in large voids Silicagel, Chemical Physics Letters,272(1-2), 91-95

\bibitem{7} Goworek, T. Ciesielski, K. Jasinska, B. Wawryszczuk, J., (1998) Positronium states in the pores of silicagel, Chemical Physics, 230(2-3), 305-315

\bibitem{8} Hohmann S., (2002), Osmotic Stress Signaling and Osmoadaptation in Yeasts,  Microbiology and molecular biology reviews, Vol. 66 No. 2, 300-372. DOI: doi: 10.1128/MMBR.66.2.300-372.2002

\bibitem{9} Hugenschmidt C., Ceeh H., (2014) The Free Volume in Dried and $ H_{2}O $-Loaded Biopolymers Studied by Positron Lifetime Measurements, J. Phys. Chem. B, 118, 9356-9360. DOI: 10.1021/jp504504p

\bibitem{10} Jasińska B, Kozioł AE, Goworek T (1996) Ortho-positronium lifetimes in nanospherical voids. J. Radioanal. Chem. 210.2, 617-623. DOI: 10.1007/BF02056403.

\bibitem{11} Jasińska B, Kozioł AE, Goworek T (1999) Void shapes and o-Ps lifetime in molecular crystals. Acta Phys. Polon. A95, 557-561

\bibitem{12} Jean Y. C.  et al., (1977) Positronium reactions in micellar systems,  J. Am. Chem. Soc. ,  99 (23) DOI:10.1021/ja00465a018

\bibitem{13} Jean Y. C.  et al., (2006) Applications of slow positrons to cancer research: Search for selectivity of positron annihilation to skin cancer, Applied Surface Science 252, 3166. DOI:10.1016/j.apsusc.2005.08.101

\bibitem{14} Kansy J (1996) Microcomputer program for analysis of positron annihilation lifetime spectra. Nucl. Instr. Methods A 374, 235-244. DOI: 10.1016/0168-9002(96)00075-7

\bibitem{15}  Liu G. , Jean Y. C.  et al., (2007) Applications of positron annihilation to dermatology and skin cancer, Phys. Stat. Sol. 4, No. 10, 3912–3915. DOI:10.1002/pssc.200675736

\bibitem{16} Moskal P et al. (2014) Patent Application No.: PCT/EP2014/068374; WO2015028604

\bibitem{17} Moskal P et al. (2014) Test of a single module of the J-PET scanner based on plastic scintillators. Nucl. Instr. and Meth. A 764,  317-321.  doi:10.1016/j.nima.2014.07.05 {\texttt{arXiv:1407.7395}} %%CITATION = ARXIV:1407.7395;%%

\bibitem{18} Moskal P et al. (2015) A novel method for the line-of-response and time-of-flight reconstruction in TOF-PET detectors based on a library of synchronized model signals. Nucl. Instr. and Meth. A 775,  54-62.  doi:10.1016/j.nima.2014.12.005; {\texttt{arXiv:1412.6963}}
  %%CITATION = ARXIV:1412.6963;%%

\bibitem{19} P. Moskal et al. (2015) Hit time and hit position reconstruction in the J-PET detector based on a library of averaged model signals Acta Phys. Pol. A127, 1495 – 1499. DOI:10.12693/AphysPolA.127.1495 {\texttt{arXiv:1502.07886}} %%CITATION = ARXIV:1502.07886;%%

\bibitem{20} Pietrzak R. et al., (2013) Influence of neoplastic therapy on the investigated blood  using positron annihilation lifetime spectroscope, Nukleonika 58, 199

\bibitem{21} Raczyński L et al. (2014) Novel method for hit-positon reconstruction using voltage signals in plastic scintillators and its application to the Positron Emission Tomography. Nucl. Instr. and Meth. A 764, 186-192.  doi:10.1016/j.nima.2014.07.032; {\texttt{arXiv:1407.8293}}
 %%CITATION = ARXIV:1407.8293;%% 

\bibitem{22} Raczyński L et al. (2015) Compressive sensing of signals generated in plastic scintillators in a novel J-PET instrument. Nucl. Instr. and Meth. A 786, 105-112.  DOI:10.1016/j.nima.2015.03.032
  
\bibitem{23}  Salgueiro, W., Somoza, A., Cabrera, O.,  Consolati, G. (2004). Porosity study on free mineral addition cement paste. Cement Concrete Res. 34(1), 91-97. DOI: 10.1016/S0008-8846(03)00258-8 
 
\bibitem{24}Tao SJ (1972) Positronium Annihilation in Molecular Substances. J. Chem. Phys. 56, 5499-5510.  DOI:10.1063/1.1677067

\bibitem{25} Wieczorek A et al. (2015) A pilot study of the novel J-PET plastic scintillator with 2-(4-styrylphenyl)benzoxazole as a wavelength shifter. Acta Phys. Pol. A 127, 1487-1490.  doi:10.12693/AphysPolA.127.1487; {\texttt{arXiv:1502.02901}}
%%CITATION = ARXIV:1502.02901;%%

  
\end{thebibliography}
\end{document}